# Interfacial current-induced torques in Pt/Co/GdOx

Satoru Emori[a], David C. Bono, and Geoffrey S. D. Beach[b]

*Department of Materials Science and Engineering, Massachusetts Institute of Technology,*

*Cambridge, MA, USA 02139*

**Abstract**

Current-driven domain wall (DW) motion is investigated in Pt/Co/GdOx nanostrips with perpendicular magnetic anisotropy. Measurements of the propagation field and the energy barrier for thermally activated DW motion reveal a large current-induced torque equivalent to an out-of-plane magnetic field of ~60 Oe per $10^{11}$ A/m$^2$. This same field-to-current scaling is shown to hold in both the slow thermally activated and fast near-flow regimes of DW motion. The current-induced torque decreases with 4 Å of Pt decorating the Co/GdOx interface and vanishes entirely with Pt replacing GdOx, suggesting that the Co/GdOx interface contributes directly to highly efficient current-driven DW motion.

E-mail:

  (a) satorue@mit.edu
  (b) gbeach@mit.edu

There has been much recent interest in using current-driven magnetic domain walls (DWs) in nanoscale ferromagnetic tracks for high-performance magnetic memory[1,2] and logic[3,4] device applications. Recent experiments show that DWs in materials with strong perpendicular magnetic anisotropy (PMA) exhibit lower critical currents for displacement than in-plane-magnetized materials, and can be driven at velocities of up to several hundred m/s by electric current alone.[5-9] Together with the narrow width and thermal stability of DWs in high-PMA materials, the enhanced efficiency of current-driven DW motion makes out-of-plane magnetized thin films well suited to achieving competitive device performance.

In thicker ferromagnetic films in which interfacial effects can be neglected, current-driven DW motion is well described by adiabatic and nonadiabatic spin-transfer torques (STTs) exerted by the conduction electron spins on the DW magnetization.[10] However, out-of-plane magnetization is usually achieved by sandwiching ultrathin ferromagnetic films between nonmagnetic high-Z metals such as Pt, Au, or Pd that generate PMA in the adjacent ferromagnet via interfacial spin-orbit coupling (SOC). As the ferromagnet is typically only a few monolayers thick, SOC at these interfaces can lead to additional current-induced torques qualitatively distinct from the usual STTs.[11-14] Miron *et al.* reported current-driven DW velocities approaching 400 m/s in Pt/Co/AlOx stacks,[6] much larger than expected from STT alone and in a direction opposite to the electron flow. They attributed these results to a large out-of-plane effective field of ~80 Oe per $10^{11}$ A/m$^2$,[15] augmented by a SOC-mediated transverse Rashba field[16] thought to stabilize the Bloch DWs such that they moved rigidly[17] rather than by precession. Evidence for a large current-induced transverse field has been independently confirmed in Pt/Co/AlOx[16,18] and Ta/CoFeB/MgO,[19] but highly efficient current-driven DW motion has not yet been reported in any metal/oxide structure beyond Pt/Co/AlOx.

It was recently argued that in addition to a transverse effective field, Rashba SOC in asymmetric structures should generate a Slonczewski-like torque[13,14] similar to that induced by perpendicular current injection in magnetic multilayers.[20] Together, these Rashba spin-orbit torques could account for the direction and high-efficiency of current-driven DW motion reported in Pt/Co/AlOx.[13] However, a Slonczewski-like torque can also arise through spin pumping from the adjacent high-Z metal via the spin Hall effect (SHE),[21,22] which was invoked to explain observations of magnetization switching by in-plane current injection.[21–23] Very recent theoretical analyses[13,14] suggest that when dissipative corrections are taken into account, the torques generated by Rashba-SOC and SHE-induced spin-pumping yield phenomenologically equivalent current-driven dynamics. It is therefore essential to isolate interfacial contributions to the current-induced torques to identify the dominant mechanisms in real materials systems.

In this Letter we examine current-driven DW motion in out-of-plane magnetized Pt/Co/GdOx films with strong PMA. We have recently shown that an electric field applied across the GdOx can control DW propagation by modulating the interfacial PMA,[24] implying that significant SOC exists at the metal/oxide interface. Strong Rashba splitting has previously been observed at the surface of oxidized Gd,[25] suggesting that Rashba-SOC might likewise manifest at the Co/GdOx interface. Here we show that an in-plane current generates a large effective out-of-plane field of ~60 Oe per $10^{11}$ A/m$^2$ that drives DWs against the electron flow direction. Decorating the Co/GdOx interface with just 4 Å of Pt diminishes the efficiency of the current-induced torque significantly, and when the GdOx is replaced by a symmetric Pt overlayer the torque vanishes entirely. These results suggest that the metal/oxide interface plays a direct role in generating this large current-induced torque.

The thin-film stack had the form Si/thermal-SiO$_2$(50)/Ta(4)/Pt(3)/Co(0.9)/GdOx(3) (numbers in parentheses indicate thicknesses in nm). The metal layers were deposited by DC magnetron sputtering under 3 mTorr of Ar at a background pressure of ~1x10$^{-7}$ Torr. The GdOx films were grown by DC reactive sputtering of a metal Gd target in an oxygen partial pressure of ~5x10$^{-5}$ Torr. The as-grown Pt/Co/GdOx films had a saturation magnetization of ~1300 emu/cm$^3$ and strong PMA with in-plane saturation field of ~8 kOe, measured using vibrating sample magnetometry.

To examine DW dynamics in these structures, 500 nm wide Pt/Co/GdOx strips with Ta(3)/Cu(100) electrodes were patterned with electron beam lithography and liftoff as shown in Fig. 1(a). DW motion was detected with a high-bandwidth scanning magneto-optical Kerr effect (MOKE) polarimeter,[26] with a focused beam spot size of ~3 μm positioned with a high-resolution sample scanning stage. DWs were initialized by the Oersted field from a 25 ns-long current pulse (~100 mA) injected through the Cu line orthogonal to the magnetic strip. The DW was then driven along the strip under combinations of out-of-plane magnetic field and an in-plane current injected along the strip as shown in Fig. 1(a). Current densities below ~1×10$^{11}$ A/m$^2$ were applied with a high-impedance DC current source. The higher current densities used in the fast DW measurements described below were injected with a voltage pulse generator turned on 50 ns before the DW initialization pulse and maintained just long enough for the DW to traverse the strip (up to 20 μs) to prevent electromigration. In all cases the injected current was monitored using an oscilloscope as shown in Fig. 1(a). The substrate temperature $T_{sub}$ was controlled using a thermoelectric module stable to ±0.1 K, and was maintained at 308 K unless otherwise noted.

We first characterized the effect of current on the DW propagation field $H_{prop}$. MOKE hysteresis loops were measured at a fixed position ~7 μm away from the DW initialization line using a triangular field sweep waveform with a frequency of 17 Hz. During each field cycle, a DW was initialized at the zero-field crossing on the rising side only using a 25 ns-long nucleation pulse as described above. The positive switching field thus corresponds to the propagation field $H_{prop}$ of the initialized DW, whereas the negative switching field is the nucleation field $H_{nuc}$ to form a reverse domain at a random location. Figure 1(b) shows magnetization switching in a Pt/Co/GdOx strip at three current densities[27] $J$ = -0.61, 0 and +0.61x$10^{11}$ A/m$^2$, each with 70 cycles averaged to account for stochasticity. $H_{nuc}$ is independent of $J$, while $H_{prop}$ increases significantly with electron flow along the field-driven propagation direction ($J > 0$), and decreases when $J$ is reversed.

The variation of $H_{prop}$ with $J$ is plotted in Fig. 1(c). Measurements were repeated on three nominally-identical strips yielding an average field-to-current ratio $\Delta H_{prop}/\Delta J$ = 57±3 Oe/$10^{11}$ A/m$^2$. As observed in other Pt/Co systems,[5,6,28,29] DW propagation is facilitated (hindered) when it is parallel (antiparallel) to the current direction. This behavior is contrary to the typical behavior under STT, which assists DW propagation in the direction of electron flow. Identical results were obtained with the opposite configuration of magnetization across the DW, realized by reversing the polarities of the driving field and initialization pulse. This demonstrates that the Oersted field from the injected current cannot play a significant role.

As shown in Fig. 1(c), when a thin Pt layer of 4 Å was inserted between the Co film and the GdOx overlayer, $\Delta H_{prop}/\Delta J$ dropped to 33±2 Oe/$10^{11}$ A/m$^2$. Moreover, no current-induced effects were observed in symmetric Pt(3)/Co(0.9)/Pt(3) strips, which were identical to Pt/Co/GdOx except for the topmost layer. These results suggest that the Co/GdOx interface plays a direct role in generating the observed large current-induced torque.

The DW propagation field depends on temperature and the timescale over which reversal is probed, and is therefore an indirect probe of thermally activated DW motion through the defect potential landscape. The DW velocity in the thermally activated regime follows an Arrhenius behavior $v=v_o exp(-E_A/kT)$, where the thermal activation energy barrier $E_A$ directly reflects the influence of the driving field and/or current on the DW dynamics. To access $E_A$ directly, we have measured thermally activated DW velocities as a function of field, current, and temperature, and used an Arrhenius analysis to unambiguously assess the influence of current on $E_A$.[30]

Average DW velocities were extracted using a time-of-flight technique as described in Ref. 26. Starting from the saturated state, a reversed driving magnetic field $H$ and current density $J$ were applied, and a reverse domain was then generated by a 25 ns current pulse in the transverse nucleation line. Time-resolved MOKE transients were then acquired as a function of position along the strip. Fig. 2(a) shows time-resolved MOKE transients (magnetization reversals) averaged over 150 cycles at several positions along the strip. The exponential tail of each averaged transient, whose breadth increases with increasing DW displacement, reflects the stochastic nature of thermally activated DW motion.[26] The average DW arrival time, taken as the time $t_{1/2}$ at which the probability of magnetization switching was 0.5, increases linearly with distance from the DW nucleation line (inset of Fig. 2(a)). These data show that DWs propagate with a uniform average velocity along the strip, governed by motion through a fine-scale disorder potential. The average DW velocity increases exponentially with driving field (Fig. 2(b)), as expected for thermally-activated propagation.

The DW velocity at four different substrate temperatures $T_{sub}$ as a function of $H$ (at $J = 0$) and $J$ (at $H = 169$ Oe) is shown in Figs. 2(c) and 2(d), respectively. The DW velocity increases by nearly an order of magnitude with $J$ parallel to the field-driven propagation direction and decreases similarly when $J$ is reversed, in agreement with the trend in $H_{prop}$ versus $J$ (Fig. 1(c)). Notably, a temperature increase of just 24 K also enhances the DW velocity by an order of magnitude, highlighting the importance of accounting for even weak Joule heating in such measurements. In Figs. 2(e) and 2(f), we have extracted $E_A$ at each driving condition, taken as the slope of $\ln(v)$ versus $T_{strip}^{-1}$, where $T_{strip} = T_{sub} + \Delta T$ and $\Delta T$ is the temperature increase due to Joule heating. $\Delta T$ was measured by comparing the strip resistance versus $T_{sub}$ at $J=0$ to the strip resistance versus $J$ at constant $T_{sub}$, giving a small correction $\Delta T = hJ^2$ with $h = 0.8$ K/[$10^{11}$ A/m$^2$]$^2$.

The data in Figs. 2(e) and 2(f) show that $E_A$ is lowered with increasing $H$ and with increasing $J$ parallel to DW motion. This strong variation of $E_A$ with $J$ in the Pt/Co/GdOx strip is in contrast with our previous results on symmetric Co/Pt multilayer strips, in which $E_A$ was insensitive to current.[30] The linear scaling of $E_A$ with field and current is consistent with the depinning regime of thermally-activated DW motion,[30,31] corresponding to the intermediate regime separating DW creep and viscous flow dynamics previously identified in this velocity range.[30] By comparing the slope of $E_A$ versus $H$ to that of $E_A$ versus $J$, we arrive at a field-to-current ratio of 67±8 Oe/$10^{11}$ A/m$^2$, in reasonable agreement with the ratio derived above from the change in $H_{prop}$ with $J$. This analysis indicates that the slope $\Delta H_{prop}/\Delta J$ provides an accurate assessment of the efficiency of current-driven DW motion.

In Fig. 3, we use the same time-of-flight technique to investigate high-speed DW motion as DW dynamics approaches the flow regime, driven by combinations of field and current.

Following the procedure in Ref. 26, the DW velocity was extracted by measuring the DW arrival time at multiple positions along the strip (Fig. 3(a)). Fig. 3(b) shows a series of DW mobility curves at several injected current densities. With current parallel to DW motion ($J < 0$), the DW mobility curves are lifted to higher velocities, approaching 200 m/s in the fastest cases. When $J$ is incorporated as part of an effective out-of-plane field $H_{eff} = H + \varepsilon J$, where $\varepsilon$ = -63 Oe/$10^{11}$ A/m$^2$, the mobility curves converge to the same dynamic scaling as shown in Fig. 3(c). Therefore, even at these high DW velocities, the influence of current can be entirely accounted for by an effective out-of-plane field with the same field-to-current ratio found from slower DW motion.

In the present experiment, random nucleation in the strip limited the maximum driving field that could be employed, preventing access to the linear flow regime of DW dynamics. The exponential increase of DW velocity with $H_{eff}$ in Fig. 3(b), together with a weak temperature dependence observed in separate measurements, indicates that DW motion at these high velocities (though approaching the viscous flow regime) is still governed by thermal activation. According to a recent finite-temperature micromagnetics study by Martinez,[32] thermally activated DW propagation at $v > 1$ m/s in a PMA nanostrip with defects occurs by DW precession. The results indicated that a DW can more readily overcome the pinning potential energy barrier by exploiting both the translational and precessional degrees of freedom. In a related study,[17] Martinez showed that a large current-induced transverse Rashba field raises the threshold driving force required for sustained DW motion by suppressing the precessional mode. Comparing Fig. 5 in Ref. 17 and Fig. 9 in Ref. 32, the current density required to move the DW at ~10 m/s increases by a factor of 4 in the presence of the Rashba field compared to the zero-Rashba field case. With the strong transverse field increasing the energy barrier for transformation between the Bloch and Néel configurations, the low-energy precessional mode is disabled and the DW can propagate only by rigid translation at higher driving currents.

If a large current-induced transverse field were present in the experiments at hand, the velocity of thermally activated DW motion at a given effective driving field $H_{eff}$ would be expected to decrease at large current densities. However, as shown in Fig. 3(c), DW dynamics remain unchanged for a fixed $H_{eff}$ even at large currents and vanishingly small applied fields. The convergence of the mobility curves to one common dynamic behavior (Fig. 3(c)) suggests that DWs in the Pt/Co/GdOx strip move by precession under all driving conditions. We thus do not observe any evidence for the existence of a transverse Rashba field sufficiently strong to qualitatively change the DW dynamics by suppressing precessional motion as indicated in Ref. 6.

In this study, we have demonstrated that a Co/GdOx interface greatly enhances the efficiency of current-driven DW motion in the ultrathin Co/Pt structure. Current injected into the Pt/Co/GdOx strip generates a strong out-of-plane effective field of ~60 Oe per $10^{11}$ A/m$^2$ that propels the DW in the direction of the current. Decoration of the Co/GdOx interface diminishes the current-induced torque, suggesting that this interface plays a direct role in the current-driven DW motion. The high efficiency of current-driven DW motion in Pt/Co/GdOx is similar to what has been observed in Pt/Co/AlOx, suggesting a common mechanism of interfacial current-induced torques in Pt/Co/oxide structures. However, we did not observe any evidence of an effective Rashba transverse field strong enough to suppress precessional DW motion.

**Acknowledgements**

This work was supported under NSF-ECCS 1128439. S.E. acknowledges support by the NSF Graduate Research Fellowship Program. The authors thank Eduardo Martinez for additional9 insights into Refs. 17 and 32. Technical assistance from Chad Kohler is gratefully acknowledged. Work was performed using instruments in the MIT Nanostructures Laboratory, the Scanning-Electron-Beam Lithography facility at the Research Laboratory of Electronics, and the Center for Materials Science and Engineering at MIT.

**References**

[1] S.S.P. Parkin, M. Hayashi, and L. Thomas, Science **320**, 190 (2008).

[2] S. Fukami, T. Suzuki, N. Ohshima, K. Nagahara, and N. Ishiwata, IEEE. T. Magn. **44**, 2539 (2008).

[3] D.A. Allwood, G. Xiong, C.C. Faulkner, D. Atkinson, D. Petit, and R.P. Cowburn, Science **309**, 1688 (2005).

[4] J. Currivan, Y. Jang, M.D. Mascaro, M.A. Baldo, and C.A. Ross, IEEE Magn. Lett. **3**, 3000104 (2012).

[5] T.A. Moore, I.M. Miron, G. Gaudin, G. Serret, S. Auffret, B. Rodmacq, A. Schuhl, S. Pizzini, J. Vogel, and M. Bonfim, Appl. Phys. Lett. **93**, 262504 (2008).

[6] I.M. Miron, T. Moore, H. Szambolics, L.D. Buda-Prejbeanu, S. Auffret, B. Rodmacq, S. Pizzini, J. Vogel, M. Bonfim, A. Schuhl, and G. Gaudin, Nat. Mater. **10**, 419 (2011).

[7] D. Chiba, G. Yamada, T. Koyama, K. Ueda, H. Tanigawa, S. Fukami, T. Suzuki, N. Ohshima, N. Ishiwata, Y. Nakatani, and T. Ono, Appl. Phys. Express **3**, 073004 (2010).

[8] T. Koyama, D. Chiba, K. Ueda, H. Tanigawa, S. Fukami, T. Suzuki, N. Ohshima, N. Ishiwata, Y. Nakatani, and T. Ono, Appl. Phys. Lett. **98**, 192509 (2011).

[9] D.-T. Ngo, K. Ikeda, and H. Awano, Appl. Phys. Express **4**, 093002 (2011).

[10] G.S.D. Beach, M. Tsoi, and J.L. Erskine, J. Magn. Magn. Mater. **320**, 1272 (2008).

[11] K. Obata and G. Tatara, Phys. Rev. B **77**, 214429 (2008).

[12] A. Manchon and S. Zhang, Phys. Rev. B **79**, 094422 (2009).

[13] K.-W. Kim, S.-M. Seo, J. Ryu, K.-J. Lee, and H.-W. Lee, Phys. Rev. B **85**, 180404 (2012).

[14] X. Wang and A. Manchon, arXiv:1111.5466 (2011).

[15] I.M. Miron, P.-J. Zermatten, G. Gaudin, S. Auffret, B. Rodmacq, and A. Schuhl, Phys. Rev. Lett. **102**, 137202 (2009).

[16] I.M. Miron, G. Gaudin, S. Auffret, B. Rodmacq, A. Schuhl, S. Pizzini, J. Vogel, and P. Gambardella, Nat. Mater. **9**, 230 (2010).

[17] E. Martinez, J. Appl. Phys. **111**, 033901 (2012).

[18] U.H. Pi, K. Won Kim, J.Y. Bae, S.C. Lee, Y.J. Cho, K.S. Kim, and S. Seo, Appl. Phys. Lett. **97**, 162507 (2010).

[19] T. Suzuki, S. Fukami, N. Ishiwata, M. Yamanouchi, S. Ikeda, N. Kasai, and H. Ohno, Appl. Phys. Lett. **98**, 142505 (2011).

[20] J.C. Slonczewski, J. Magn.Magn. Mater. **159**, L1 (1996).

[21] L. Liu, O.J. Lee, T.J. Gudmundsen, D.C. Ralph, and R.A. Buhrman, arXiv:1110.6846 (2011).

[22] L. Liu, C.-F. Pai, Y. Li, H.W. Tseng, D.C. Ralph, and R.A. Buhrman, Science **336**, 555 (2012).

[23] I.M. Miron, K. Garello, G. Gaudin, P.-J. Zermatten, M.V. Costache, S. Auffret, S. Bandiera, B. Rodmacq, A. Schuhl, and P. Gambardella, Nature **476**, 189 (2011).

[24] U. Bauer, S. Emori, and G.S.D. Beach, Appl. Phys. Lett. **100**, 192408 (2012).

[25] O. Krupin, G. Bihlmayer, K. Starke, S. Gorovikov, J.E. Prieto, K. Döbrich, S. Blügel, and G. Kaindl, Phys. Rev. B **71**, 201403 (2005).

[26] S. Emori, D.C. Bono, and G.S.D. Beach, J. App. Phys. **111**, 07D304 (2012).

[27] The reported current density is the average value through the Pt and Co layers, assuming a current distribution in the Ta/Pt/Co stack based on the bulk resistivities of the individual layers.

[28] K.-J. Kim, J.-C. Lee, S.-J. Yun, G.-H. Gim, K.-S. Lee, S.-B. Choe, and K.-H. Shin, Appl. Phys. Express **3**, 083001 (2010).

[29] J.-C. Lee, K.-J. Kim, J. Ryu, K.-W. Moon, S.-J. Yun, G.-H. Gim, K.-S. Lee, K.-H. Shin, H.-W. Lee, and S.-B. Choe, Phys. Rev. Lett. **107**, 067201 (2011).

[30] S. Emori and G.S.D. Beach, J. Phys. Condens. Mat. **24**, 024214 (2012).

[31] J. Ferré, *Spin Dynamics in Confined Magnetic Structures I*, edited by B. Hillebrands and K. Ounadjela (Springer, Berlin, 2002), pp. 127–165.

[32] E. Martinez, J. Phys. Condens. Mat. **24**, 024206 (2012).

**Figure Captions**

FIG. 1. (Color online) (a) Micrograph of a 500-nm wide Pt/Co/GdOx strip and measurement schematic. With $J < 0$, the current is in the same direction as (and electron flow opposes) field-driven DW motion. (b) Hysteresis loops showing the DW propagation field changing and nucleation field invariant with injected DC current densities ($J$ = -0.61, 0, and +0.61x10$^{11}$ A/m$^2$). Note the breaks in the horizontal scale to show details. (c) Plot of the DW propagation field change $\Delta H_{prop}$ against the injected DC current density in Pt/Co/GdOx, Pt/Co/Pt/GdOx, and Pt/Co/Pt strips (whose zero-current propagation fields are 170, 250, and 160 Oe, respectively).

FIG. 2. (Color online) (a) Averaged DW transients probed at several positions at $T_{sub}$ = 324 K, $H$ = 169 Oe, $J$ = +1.05x10$^{11}$ A/m$^2$. The inset is a plot of average DW arrival time $t_{1/2}$ against probed position, with the linear fit indicating a uniform average velocity. (b) Purely field-driven DW velocity spanning more than 5 decades at $T_{sub}$ = 308 K. (c, d) DW velocity at several substrate temperatures versus applied field with $J$ = 0 (c) and versus current density with $H$ = 169 Oe (d). (e, f) Activation energy versus applied field (e) and versus current density. (f)

FIG. 3. (Color online) (a) Averaged DW transients at three positions at $T_{sub}$ = 308 K, $H$ = 130 Oe, $J$ = -6.5×10$^{11}$ A/m$^2$. The inset is a plot of average DW arrival time $t_{1/2}$ against probed position, with the linear fit to extract the DW velocity. (b) DW velocity versus applied field at $T_{sub}$ = 308 K at several different current densities. (c) DW velocity plotted against effective magnetic field $H_{eff} = H+\varepsilon J$.

**FIG. 1**

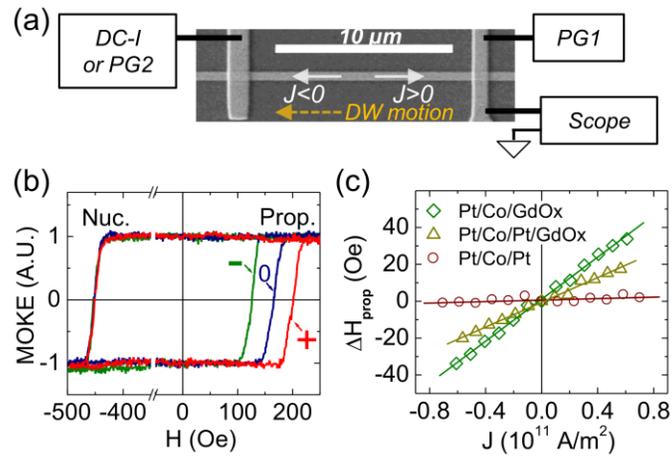

**FIG. 2**

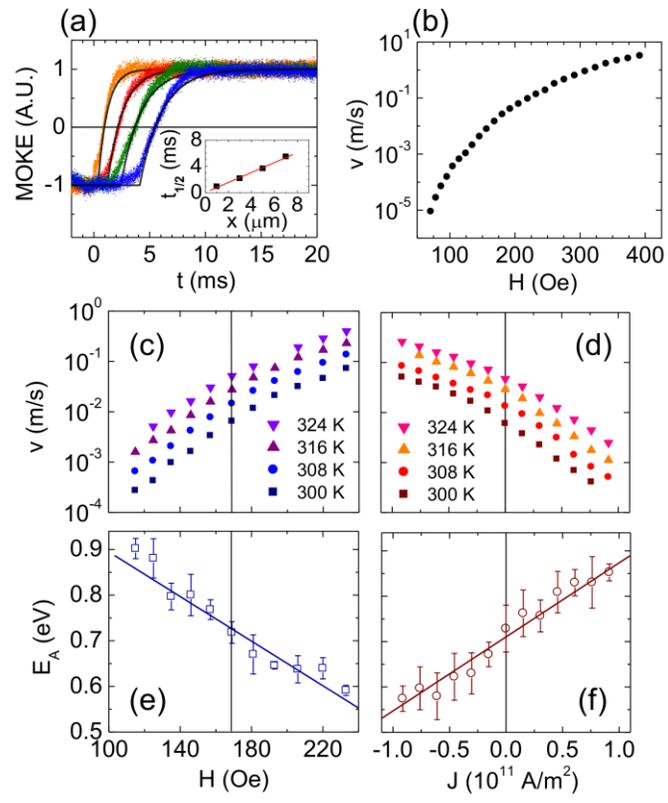

**FIG. 3**

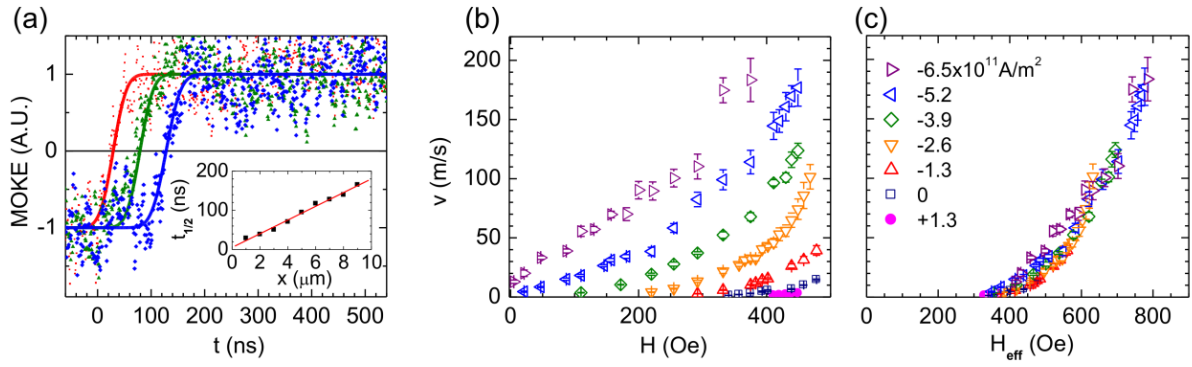